\documentclass[doublecol]{epl2} 
\usepackage{amsmath}
\usepackage{amsfonts}
\usepackage{mathtools}

\DeclarePairedDelimiter\ket{\lvert}{\rangle}
\DeclarePairedDelimiterX\braket[2]{\langle}{\rangle}{#1 \delimsize\vert #2}
\usepackage[svgnames]{xcolor}
\usepackage[colorlinks=true, linkcolor=Maroon, urlcolor=Blue]{hyperref} 
\usepackage{color}
\usepackage{microtype}
\DisableLigatures{encoding = *, family = *}
\usepackage[figuresleft]{rotating}
\usepackage{relsize}
\usepackage[rightcaption]{sidecap}
\usepackage[figuresleft]{rotating}
\usepackage{float}
\usepackage{subfigure}
\usepackage{graphicx}
\hypersetup{
  colorlinks,
  citecolor=MidnightBlue,
  linkcolor=DarkRed,
  urlcolor=Blue}

\title{Quantized Josephson phase battery}

\author{Subhajit Pal\inst{1} \and Colin Benjamin\inst{1}}
\institute{ School of Physical Sciences, National Institute of Science Education \& Research, HBNI, Jatni-752050,\ India                    
  \inst{1} 
}
\pacs{03.67.-a}{ Josephson junction}
\pacs{02.50.Le}{Battery}
\pacs{03.67.Ac}{Anomalous phase}
\abstract{{A ferromagnetic Josephson junction with a spin-flipper (magnetic impurity) sandwiched in-between acts as a phase battery that can store quantized amounts of superconducting phase difference $\varphi_{0}$ in the ground state of the junction. Moreover, for such $\varphi_{0}$-Josephson junction anomalous Josephson current appears at zero phase difference. We study the properties of this quantum spin-flip scattering induced anomalous Josephson current, especially its tun-ability via misorientation angle between two Ferromagnets.}}
\begin{document}
\maketitle
\section{Introduction}
Josephson Free energy, in general, is minimum when phase difference across the Josephson junction is either zero for $0$-junction or $\pi$ for a $\pi$-junction\cite{golu}. In such junctions, Josephson super-current vanishes when phase difference between two superconductors is zero as current-phase relation is sinusoidal $I(\varphi)=I_{c}\sin(\varphi)$, with $\varphi$ being the phase difference across superconductors and $|I_{c}|$ is the maximum supercurrent flowing through the junction\cite{golu}. However, Josephson Free energy can sometimes be minimum at a phase difference $\varphi_{0}$ ($\neq0$ or $\pi$). The current-phase relation in such $\varphi_{0}$- Josephson junction's\cite{buz,buzd,tan} satisfies
$I(\varphi)=I_{c}\sin(\varphi+\varphi_{0})$, i.e., there is a phase shift $\varphi_{0}$ in the conventional current-phase relation. This suggests that Josephson current can flow even at zero phase difference ($\varphi=0$) between two superconducting electrodes\cite{sil,liu,Liu,GC,VB}. This effect is known as anomalous Josephson effect (AJE), and $I_{an}=I(0)=I_{c}\sin(\varphi_{0})$ is referred to as anomalous Josephson current. 

The physics behind anomalous Josephson effect is naturally linked with breaking of some symmetries of the system\cite{chan,ARA}. One of them is time reversal symmetry and it implies $I(-\varphi)=-I(\varphi)$, which results in $I(\varphi=0)$ being zero. So, when system preserves time reversal symmetry there is no anomalous current in the device. However, breaking time reversal symmetry is a necessary but not sufficient condition to produce anomalous Josephson current at $\varphi=0$. In junctions with ferromagnetic coupling time reversal symmetry is broken, but there is no anomalous Josephson current\cite{BUZ,BER}. This implies some other symmetry is present in the system which prevents the appearance of anomalous Josephson current at $\varphi=0$. This symmetry is called chiral symmetry\cite{kriv} which ensures that at $\varphi=0$ the tunneling amplitude relating electron tunneling from left superconductor to right superconductor is exactly same as the one related to tunneling in reverse, i.e., from right to left superconductor. These leftward and rightward tunneling processes cancel each other, leading to vanishing current flow at $\varphi=0$. Thus, to have anomalous Josephson current at $\varphi=0$, one needs to break both symmetries. Different ways have been suggested earlier to break these symmetries and generate anomalous Josephson current. These include Josephson junctions with conventional s-wave superconductors in presence of both spin-orbit interaction and Zeeman field\cite{yoko,nes}, ferromagnetic Josephson junctions with non-coplanar magnetizations\cite{Liu}, SNS junctions with s-wave superconductors where N region is a magnetic normal metal\cite{YA,ME,FK}, a quantum dot\cite{AZ,LD} or a quantum point contact\cite{AAR,AA}. Further, anomalous Josephson effect can also be found in systems with unconventional superconductors\cite{lu,SK,RG,SY,YT}. Experimentally, $\varphi_{0}$ phase shift has been recently predicted in a Josephson junction based nanowire quantum dot\cite{szo}. More interestingly some Josephson junctions reveal the remarkable feature that the phase shift $\varphi_{0}$ is accompanied by a direction dependent critical current $(I_{c+}\neq I_{c-})$, where $I_{c+}$ and $I_{c-}$ are the absolute values of maximum and minimum Josephson current respectively.

In this work we study anomalous Josephson effect and the direction dependent critical current in a junction consisting of two Ferromagnet's with mis-aligned magnetizations and a spin-flipper sandwiched between two $s$-wave superconductors. This system acts as a quantized phase battery which can supply anomalous current even at zero phase difference. The main advantage of our system over all other proposals involving anomalous Josephson current is that our system can store quantized amounts of phase $\varphi_{0}$ in the ground state of junction. The reason we are interested in $\varphi_{0}$ Josephson junction is because of the manifold applications of such junctions in phase qubits\cite{CP}, superconducting computer memory components\cite{EC}, superconducting phase batteries\cite{TO} and in rectifiers\cite{AA}. Our manuscript is organized as follows: in the next section we present the model Hamiltonian and explain the steps necessary to calculate the Josephson current. Following this we discuss our results for the Andreev bound states and anomalous Josephson currents, symmetries broken in our system when anomalous current flows through the junction and plot the quantized anomalous phase. {We next discuss effect of change of temperature on anomalous Josephson effect.} Finally, we conclude with an experimental realization and summary of our work wherein we provide a table on condition necessary for seeing Anomalous Josephson effect in our setup.
\section{Theory}
\subsection{\textbf{Hamiltonian}}
Our set-up is depicted in Fig.~1, it shows a spin-flipper at $x=0$ and two superconductors- one to left $x<-a/2$ and another at right $x>a/2$. There are two Ferromagnet's in between at $-a/2<x<0$ and $0<x<a/2$. The magnetization vectors of the two Ferromagnet's make an angle $\theta$ with each other. We take the superconducting gap of the form {$\Delta=\Delta_{0}(T)[e^{i\varphi_{L}}\Theta(-x-a/2)+e^{i\varphi_{R}}\Theta(x-a/2)]$, where $\Delta_{0}(T)$ is temperature dependent and it follows that $\Delta_{0}(T)= \Delta_{0}\tanh(1.74\sqrt{(T_{c}/T-1)})$}, where $T_{c}$- the superconducting critical temperature\cite{annu} {for a widely used s-wave superconductor like lead is $7.2$K}, $\varphi_{L}$ and $\varphi_{R}$ being superconducting phases for left and right superconductors respectively.
\begin{figure}[ht]
\begin{center} 
\includegraphics[width=0.48\textwidth]{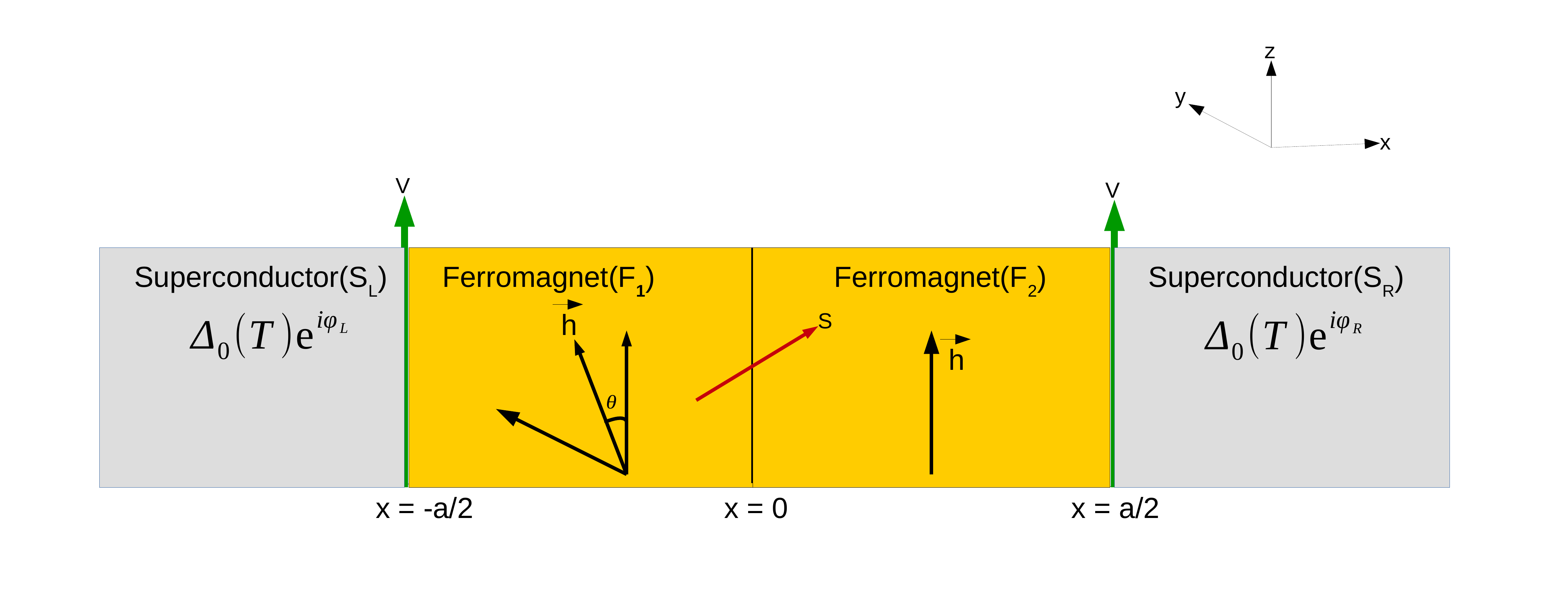}
\caption{\small \sl Josephson junction with two Ferromagnet's and a spin-flipper (spin $S$, magnetic moment $m'$) at $x=0$ sandwiched between two s-wave superconductors.}
\end{center}
\end{figure}
{The Bogoliubov-de Gennes equation for our junction is\cite{LINDER}}-
\begin{eqnarray}
\begin{pmatrix}
H\hat{I} & i\Delta \hat{\sigma}_{y}\\
-i\Delta^{*} \hat{\sigma}_{y} & -H^{*}\hat{I}
\end{pmatrix} \psi(x)& =& E \psi(x), \mbox{}
\label{eqq}
\end{eqnarray}
where $H=p^2/2m^{\star}+V[\delta(x+a/2)+\delta(x-a/2)]-J_{0}\delta(x)\vec s.\vec S-\vec{h}.\hat{\sigma}[\Theta(x+a/2)+\Theta(a/2-x)]-E_{F}$, with $p^2/2m^{\star}$ being the kinetic energy of electron with mass $m^{\star}$, $V$ denotes the strength of $\delta$ potentials at the interfaces between Ferromagnet's and Superconductor, $J_{0}$ denotes strength of exchange coupling between electron/hole with spin $\vec{s}$ and spin-flipper\cite{joseph,Fra} with spin $\vec{S}$. $\psi(x)$ defines a four-component spinor, while $E_{F}$ is the Fermi energy, $\hat{\sigma}$'s are Pauli spin matrices and $\hat{I}$ is $2\times2$ identity matrix. The magnetization vector ($\vec{h}$) of left ferromagnetic layer ($F_{1}$) is at an angle $\theta$ with $z$ axis in the $y-z$ plane, while that of right ferromagnetic layer ($F_{2}$) is fixed along the $z$ axis. Thus, $\vec{h}.\hat{\sigma}=h\sin \theta\hat{\sigma}_{y}+h\cos \theta\hat{\sigma}_{z}$\cite{Halter}. {In the rest of the paper, we use the dimensionless parameters $J=\frac{m^{\star}J_{0}}{\hbar^2k_{F}}$ as a measure of strength of exchange interaction\cite{AJP} and $Z=\frac{m^{\star}V}{\hbar^2 k_{F}}$ as a measure of interface transparency\cite{BTK}}. The wavefunctions, boundary conditions of our system as depicted in Fig.~1 and the calculations of Andreev bound states are mentioned in supplementary material.
\subsection{\textbf{Anomalous Josephson current}}
On solving the boundary conditions, see supplementary material, we get the Andreev bound state energy spectrum\cite{Been} $E_{i}(i=1,...,8)=\pm\varepsilon_{l}(l=1,...,4)$. From Andreev bound state energies the Free energy of our junction can be calculated\cite{annu} as-
\begin{equation}
F=-\frac{1}{\beta}\ln\Big[\prod_{i}(1+e^{-\beta E_{i}})\Big]= -\frac{2}{\beta}\sum_{l=1}^{4}\ln\Big[2\cosh\Big(\frac{\beta \varepsilon_{l}}{2}\Big)\Big]
\label{ff}
\end{equation}
where  $E_{i}(i=1,...,8)=\pm\varepsilon_{l}(l=1,...,4)$ defines Andreev bound state energy spectrum. In short junction limit considered in this paper, total Josephson current can be determined by bound state contribution only. The Josephson current at finite temperature is defined as the derivative of the Free energy $F$ of our system with respect to the phase difference $\varphi$ between left and right superconductors\cite{deG},
\begin{equation}
I=\frac{2e}{\hbar}\frac{\partial F}{\partial \varphi}=-\frac{2e}{\hbar}\sum_{l=1}^{4}\tanh\Big(\frac{\beta \varepsilon_{l}}{2}\Big)\frac{\partial \varepsilon_{l}}{\partial \varphi},
\label{ff1}
\end{equation}
herein $e$ is the charge of electron. Eq.~\ref{ff1} is the main working formula of our paper. Using Eq.~\ref{ff1} we calculate anomalous Josephson current, as- $I_{an}=I(\varphi=0)$, {absolute value of maximum Josephson current, as- $I_{c+}=|\max I(\varphi)|$ and absolute value of minimum Josephson current, as- $I_{c-}=|\min I(\varphi)|$}. In case interfaces are completely transparent, i.e., $Z=0$, we have-
\begin{align}
\begin{split}
I_{an}={}&\frac{2e\Delta_{0}^{2}(T)}{\hbar}\Big(\tanh\Big(\frac{\beta A_{1}}{2}\Big){A_{1}^{\prime}}+\tanh\Big(\frac{\beta A_{2}}{2}\Big){A_{2}^{\prime}}\\{}&+\tanh\Big(\frac{\beta A_{3}}{2}\Big){A_{3}^{\prime}}+\tanh\Big(\frac{\beta A_{4}}{2}\Big){A_{4}^{\prime}}\Big)
\label{an1}
\end{split}
\end{align}
wherein $A_{1}$, $A_{2}$, $A_{3}$, $A_{4}$, $A_{1}^{\prime}$, $A_{2}^{\prime}$, $A_{3}^{\prime}$ and $A_{4}^{\prime}$ are expressions that depend on exchange interaction ($J$), magnetization of the Ferromagnet's, spin ($S$) and magnetic moment ($m'$) of spin-flipper, phase ($k_{F}a$) accumulated in Ferromagnet's and spin-flip probability ($f_{2}$). The explicit forms for $A_{k}$'s and $A_{k}^{\prime}$'s ($k=1,2,3,4$) are given in Appendix of supplementary material. In Appendix of supplementary material we show that for no flip ($f_{2}=0$) or absence of spin-flipper ($J=0$) or $\theta=0$ (magnetizations of Ferromagnet's are aligned), anomalous Josephson current vanishes.
\begin{figure*}[ht]
\begin{center} 
\includegraphics[width=0.85\textwidth]{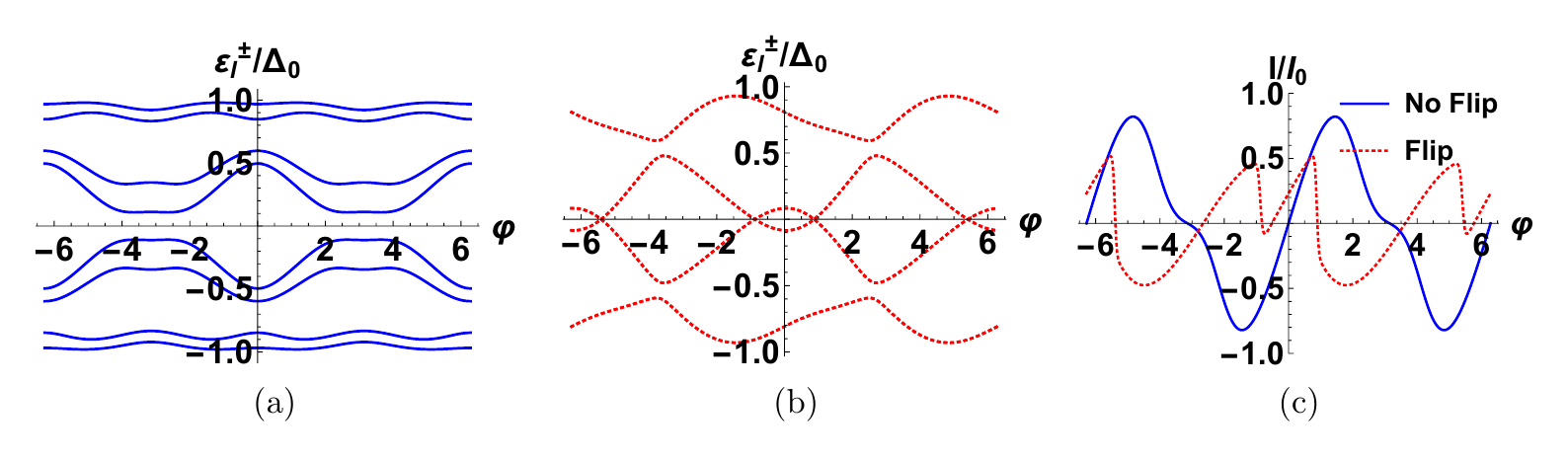}
\caption{\small \sl Andreev bound state energies as a function of phase difference ($\varphi$) (a) for no flip, (b) for flip. (c) Josephson current as a function of phase difference ($\varphi$). Parameters are $\Delta_{0}=1meV$, $J=1$, $h=0.5E_{F}$, $I_{0}=e\Delta_{0}/\hbar$, $T/T_{c}=0.01$, $Z=0$, $k_{F}a=1.2\pi$, $\theta=\pi/2$, for flip: $S=3/2$, $m'=-1/2$, $f_{2}=2$, and for no flip:   $S=3/2$, $m'=3/2$, $f_{2}=0$.}
\end{center}
\end{figure*}
\section{Results}
\subsection{\textbf{Anomalous Josephson current and Andreev bound states}}
In this subsection, we first show results for Andreev bound states and Josephson currents of our junction. In Fig.~2 we plot Andreev bound states and Josephson current as a function of phase difference $\varphi$ between two superconductors for both no flip and spin flip cases.
In Fig.~2(a) we deal with the no flip case, i.e., $f_{2}=0$ (see section I.~C in supplementary material), in this case $S=m'$, i.e., $S=3/2$, $m'=3/2$, { where $m'$ is the spin magnetic moment in $z-$ direction for the spin-flipper defined as $S_{z} \phi_{m'}^{S}=m' \phi_{m'}^{S}$, $S_{z}$ being spin-operator in z-direction acting on spin flipper wave function $\phi_{m'}^S$}. There is no possibility for spin-flipper to flip its own spin while interacting with an electron/hole if $S=m'$, for an explanation see section I. C of supplementary material. However, there is a finite probability for the spin of electron or hole to flip due to the misalignment in the magnetization of the Ferromagnet's. We see that there are four positive and four negative Andreev levels. In junction where time reversal symmetry is not broken, Andreev bound states $\varepsilon_{l}(\varphi)$ are invariant with respect to inversion of phase difference $\varphi$, i.e., $\varepsilon_{l}(-\varphi)=\varepsilon_{l}(\varphi)$. As a result, in Fig.~2(c), for no flip case Josephson current satisfies $I(-\varphi)=-I(\varphi)$ and there is no current flowing through the junction when phase difference $\varphi$ between two superconductors is zero. Thus, absolute value of maximum Josephson current, $I_{c+}$ is identical to absolute value of minimum Josephson current, $I_{c-}$. In Fig.~2(b) we deal with spin-flip case, i.e., $f_{2}\neq0$ (see section I.~C in supplementary material), for this case $S\neq m'$, i.e., $S=3/2$, $m'=-1/2$, for spin-flipper. Thus, there is finite probability for spin-flipper to flip its own spin when interacting with electron/hole. We see that for $m'=-1/2$, Andreev levels are doubly degenerate and phase inversion symmetry, i.e.,  $\varepsilon_{l}(-\varphi)=\varepsilon_{l}(\varphi)$ is broken. As a result, anomalous Josephson current flows, i.e., $I(-\varphi)\neq-I(\varphi)$ for spin-flip process in Fig.~2(c), where not only the anomalous current $I(\varphi=0)\neq0$, but also difference between the absolute value of maximum and absolute value of minimum Josephson currents, $I_{c+}\neq I_{c-}$ is finite. 

{In Fig.~3 we show the effects of exchange interaction ($J$) of spin-flipper, magnetization of Ferromagnet's ($h$), interface transparency ($Z$) and mis-orientation angle ($\theta$) between two ferromagnetic layers on the anomalous Josephson current. In Fig.~3(a) we plot anomalous Josephson current as a function of exchange interaction $J$ of spin flipper for different spin-flip probabilities.
\begin{figure}[ht]
\begin{center} 
\includegraphics[width=0.49\textwidth]{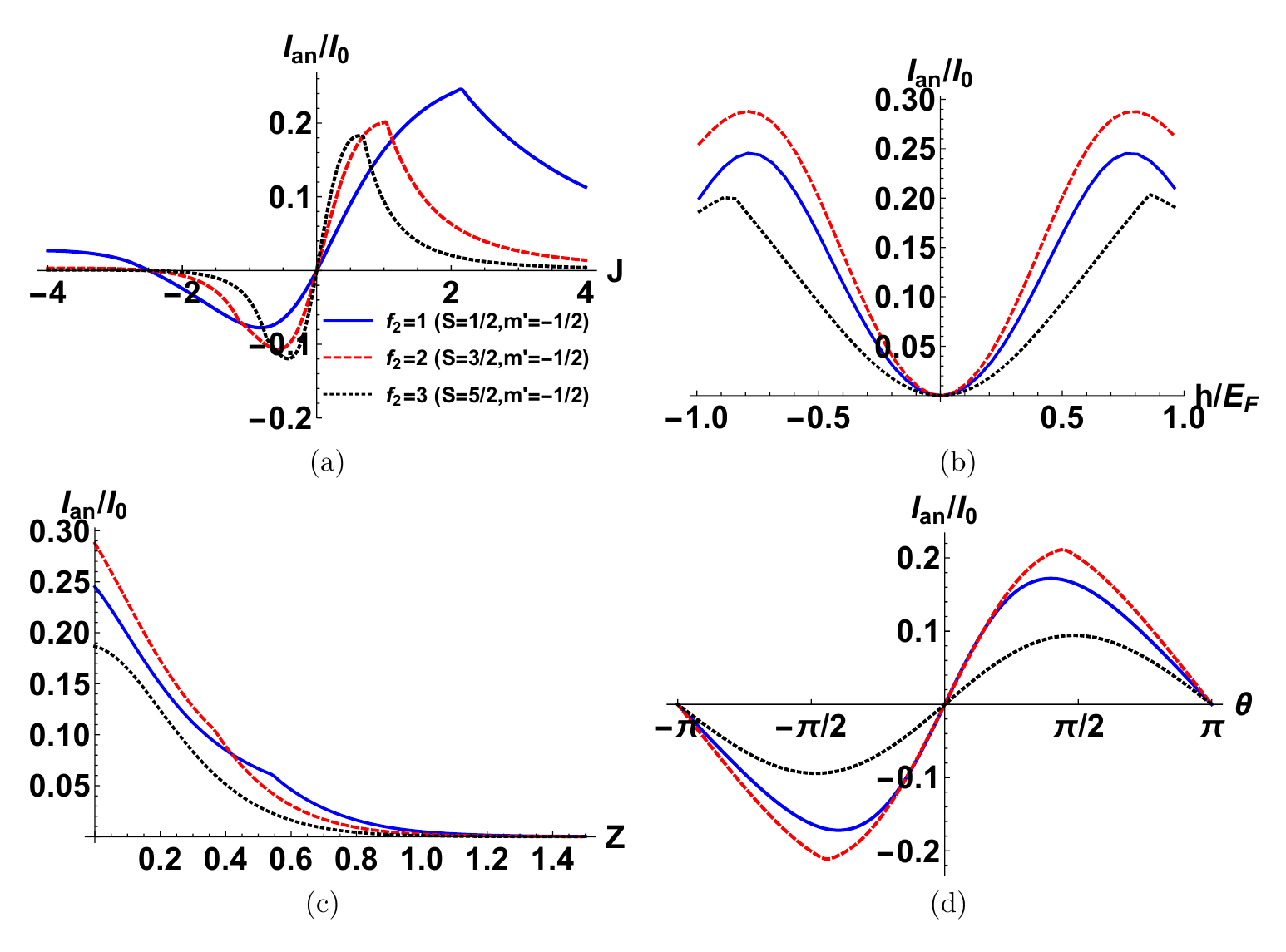}
\caption{\small \sl (a) Anomalous Josephson current as a function of exchange interaction $J$ of spin flipper, (b) Anomalous Josephson current as a function of magnetization ($h$) of the Ferromagnet's, (c) Anomalous Josephson current as a function of the interface barrier strength ($Z$), (d) Anomalous Josephson current as a function of mis-orientation angle ($\theta$) between two Ferromagnets' for different spin flip probabilities of spin flipper. Parameters are $\Delta_{0}=1meV$, $I_{0}=e\Delta_{0}/\hbar$, $T/T_{c}=0.01$, $J=1$ (for (b), (c) and (d)), $h=0.5E_{F}$ (for (a), (d)), $h=0.8E_{F}$ (for (c)), $k_{F}a=\pi$, $Z=0$ (for (a), (b) and (d)), $\theta=\pi/2$ (for (a), (b) and (c)).}
\end{center}
\end{figure}
We see that for ferromagnetic coupling ($J>0$) there is no sign change of anomalous Josephson current with change in $J$. However, for anti-ferromagnetic coupling ($J<0$) there is a sign change in $I_{an}$ as $J$ changes from $J=0$ to $J=-4$, implying tun-ability of the sign of anomalous Josephson current via the exchange interaction of spin flipper. We also see that anomalous Josephson current is asymmetric with respect to $J$. Further, the maximum value of $I_{an}$ decreases with increase of spin flip probability of spin flipper. In Fig.~3(b) we plot anomalous Josephson current as a function of magnetization ($h$) of the ferromagnetic layers. In contrast to Fig.~3(a), anomalous Josephson current is symmetric with respect to magnetization $h$ of the Ferromagnet's. In Fig.~3(c) we plot $I_{an}$ as a function of interface barrier strength ($Z$). We see that there is no sign change of $I_{an}$ with increase of interface barrier strength $Z$. Further, anomalous Josephson current is almost zero in the tunneling regime. It is also evident from Fig.~3(b) and Fig.~3(c) that maximum of $I_{an}$ decreases for large values of spin flip probability. In Fig.~3(d) $I_{an}$ is plotted as a function of mis-orientation angle ($\theta$) between two ferromagnetic layers. We see that the magnitude of anomalous current decreases with increasing spin-flip probability.  
Further, one can see that sign of anomalous Josephson current can be tuned via the mis-orientation angle $\theta$ between two Ferromagnet's. Anomalous current is periodic as function of mis-orientation angle with period $2\pi$. From Fig.~3(d) we also see that when the magnetic moments of the Ferromagnet's are aligned parallel or anti-parallel ($\theta=0$ or $\theta=\pi$), anomalous Josephson current vanishes even when spin flipper flips its spin. In supplementary material (section I.~D) we plot absolute value of the anomalous Josephson current as function of the mis-orientation angle ($\theta$) between two Ferromagnets for same parameters as in Fig.~3(d).} 
\subsection{\textbf{Quantized anomalous phase}}
We have seen the results of Andreev bound states and Anomalous Josephson current in Figs. 2 and 3. Now we discuss the results for anomalous phase $\varphi_{0}$ (see section I.~D in supplementary material for the method to calculate $\varphi_{0}$). In Fig.~4 we plot anomalous phase $\varphi_{0}$ as a function of exchange interaction $J$ of spin flipper and magnetization $h$ of the Ferromagnet's.
\begin{figure*}[ht]
\begin{center} 
\includegraphics[width=0.85\textwidth]{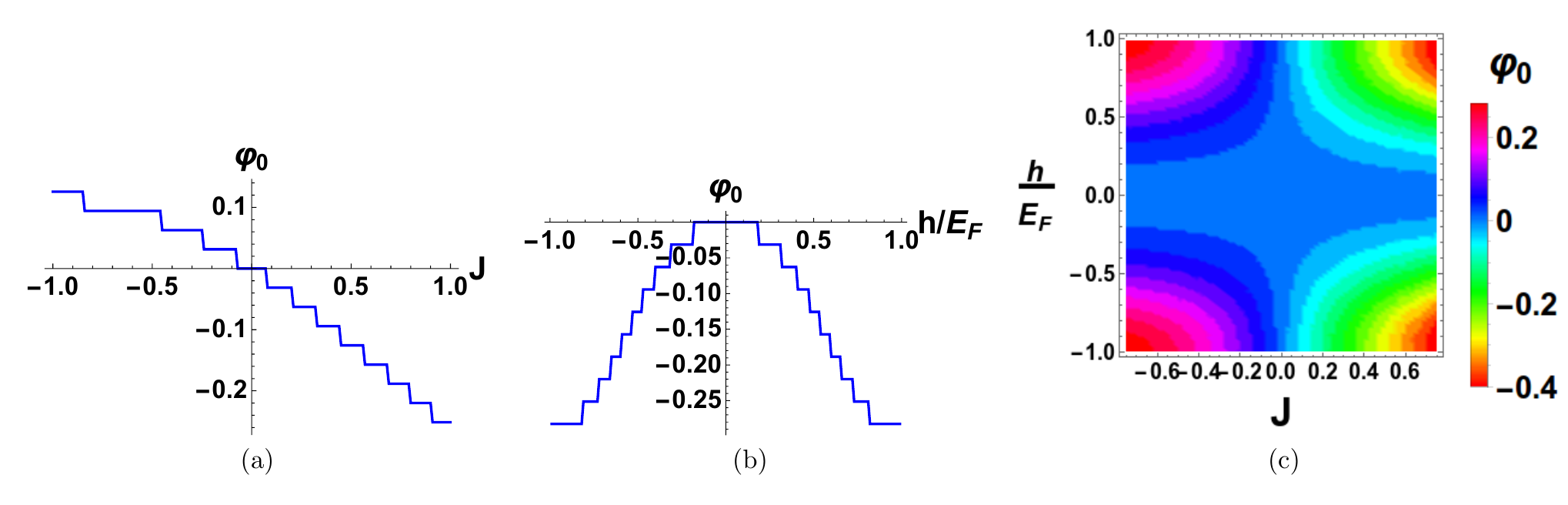}
\caption{\small \sl (a) Phase difference $\varphi_{0}$ as a function of exchange interaction ($J$) of spin flipper.   
(b) Phase difference $\varphi_{0}$ as a function of magnetization ($h$) of Ferromagnet's. (c) Density plot of $\varphi_{0}$ as a function of exchange interaction ($J$) of spin flipper and magnetization ($h$) of Ferromagnet's. Parameters are $\Delta_{0}=1meV$, $T/T_{c}=0.01$, $S=1/2$, $m'=-1/2$, $h=0.5E_{F}$ (for (a)), $k_{F}a=\pi$, $\theta=\pi/2$, $Z=0$, $J=0.5$ (for (b)).}
\end{center}
\end{figure*}
In Fig.~4(a) we see \textbf{\textquotedblleft quantized\textquotedblright} steps in the anomalous phase $\varphi_{0}$ which are of exactly same magnitude ($\pi/100$ radians) although the width linearly decreases as one goes from anti-ferromagnetic to ferromagnetic coupling. In our work anomalous current is always accompanied by quantized anomalous phase. We never see anomalous current with non quantized anomalous phase. In Fig.~4(b) anomalous phase $\varphi_{0}$ is shown as a function of the normalized magnetization $h/E_{F}$ of the Ferromagnet's. Similar to Fig.~4(a), we also see quantized steps in anomalous phase $\varphi_{0}$ which are again exactly of same magnitude ($\pi/100$ radians) although the width initially decreases and then increases as one changes $h/E_{F}$ from $-0.99$ to $0$ and then from $0$ to $0.99$. The quantized step magnitude or height remains same for different values of spin, magnetic moment and different spin flip probability of spin flipper. Quantized behavior of $\varphi_{0}$ is also shown in Fig.~4(c), where we show density plot of $\varphi_{0}$ as a function of $J$ and $h$. It is also evident from Fig.~4(c) larger values of exchange interaction and magnetization correspond to increasing magnitudes of anomalous phase $\varphi_{0}$. In Ref.~\cite{yoko}, anomalous Josephson current is seen in a semiconducting nanowire based junction in presence of both spin-orbit interaction (SO) and Zeeman field, which is equivalent to what we see for finite spin flip probability in our system without any need for spin orbit scattering and/or Zeeman fields. However, in Ref.~\cite{yoko} anomalous phase $\varphi_{0}$ changes continuously with change in magnetic field and is not quantized  in contrast to what is shown in this work.

Finally, in Fig.~5 we perform a similar analysis for the asymmetry of the critical current, defined as\cite{FO} $\aleph=(I_{c+}-I_{c-})/(I_{c+}+I_{c-})$. In Fig.~5(a) we plot $\aleph$ as a function of the exchange interaction $J$ and see that maximum value of $\aleph$ almost remains same for different spin flip probabilities. Further, it is also evident from Fig.~5(a), the sign of $\aleph$ can be tuned via $J$, and $\aleph$ is asymmetric with respect to $J$. Figure 5(b) shows the asymmetry $\aleph$ as a function of magnetization $h$ of Ferromagnet's. We see that in contrast to Fig.~5(a), maximum in $\aleph$ is different for different spin flip probabilities. Further, $\aleph$ is symmetric with respect to $h/E_{F}$. In Fig.~5(c) we show a density plot for asymmetry in critical current ($\aleph$) as a function of exchange interaction $J$ of spin flipper and magnetization $h$ of Ferromagnet's. We find maximum value of $\aleph\simeq0.16$ with $\aleph$ changing sign from anti-ferromagnetic to ferromagnetic coupling. Asymmetry of the critical current is also calculated in Ref.~\cite{FO} as a function of the spin orbit interaction and the magnetization. But, in contrast to our case, $\aleph$ is larger for larger values of spin orbit interaction and magnetization.
\begin{figure*}[ht]
\begin{center} 
\includegraphics[width=0.85\textwidth]{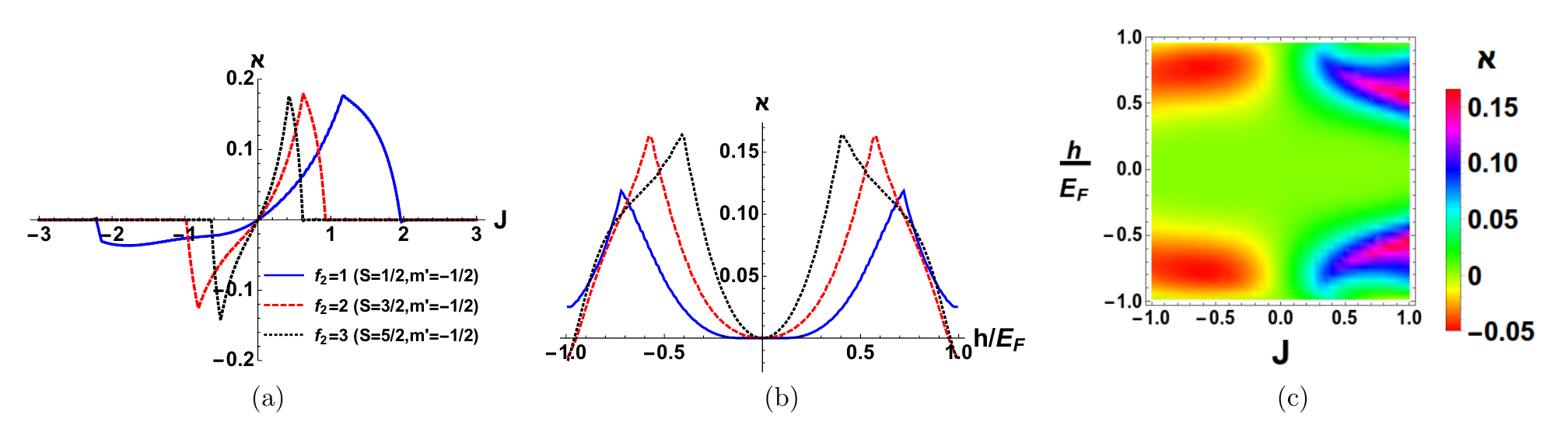}
\caption{\small \sl (a) Asymmetry of critical current as a function of exchange interaction ($J$) of spin flipper. (b) Asymmetry of critical current as a function of magnetization ($h$) of Ferromagnet's. (c) Asymmetry of the critical current as a function of exchange interaction ($J$) of spin-flipper and magnetization ($h$) of Ferromagnet's. Parameters are $\Delta_{0}=1meV$, $T/T_{c}=0.01$, $f_{2}=1$ ($S=1/2$, $m'=-1/2$) (for (c)), $h=0.5E_{F}$ (for (a)), $k_{F}a=\pi$, $\theta=\pi/2$, $Z=0$, $J=0.5$ (for (b)).}
\end{center}
\end{figure*}
\begin{figure*}[ht]
\begin{center}
\includegraphics[width=0.9\linewidth]{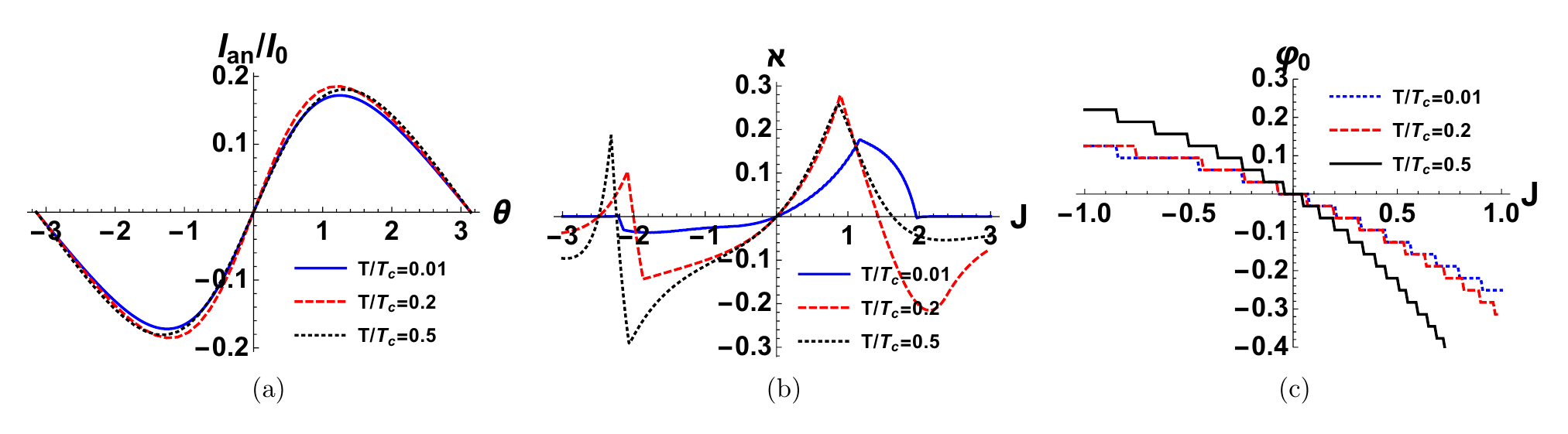}
\caption{\small \sl (a) Anomalous Josephson current as a function of mis-orientation angle ($\theta$) between two Ferromagnets' for different values of temperature, (b)  Asymmetry of the critical current for different values of temperature calculated as a function of exchange interaction ($J$) of spin flipper, (c) Phase difference $\varphi_{0}$ as a function of exchange interaction ($J$) of spin flipper for different values of temperature. Parameters are $\Delta_{0}=1meV$, $J=1$ (for (a)), $h=0.5E_{F}$ (for (a), (b) and (c)), $I_{0}=e\Delta_{0}/\hbar$, $f_{2}=1$ ($S=1/2$, $m'=-1/2$), $Z=0$, $k_{F}a=\pi$, $\theta=\pi/2$ (for (b) and (c))}
\end{center}
\end{figure*}
\subsection{\textbf{Reasons for existence of Anomalous Josephson effect}}

{\textbf{Explaining quantum spin flip scattering-}}
The extremely important role played by the spin flipper entails a detailed analysis of this process. The Josephson current flowing through either ferromagnetic layer ($F_{1}$ or $F_2$) is spin-polarized in direction of magnetization of that ferromagnetic layer. Subsequently when this spin polarized Josephson current state, denoted by a macroscopic wave-function $\sim|\Psi_{S_{P}}|e^{i\varphi_{P}}\approx (u\hspace{2pt}0\hspace{2pt}0\hspace{2pt}v)^{T} e^{i\varphi_{P}}$ (where $P=L$ or $R$, i.e., Left or Right superconductor), interacts with the spin flipper, there is finite probability for mutual spin-flip. This of course is a probability not a certainty, since the interaction of spin polarized Josephson current with spin-flipper is quantum in nature, see Eq.~(7) of supplementary material. Thus, the combined state of spin polarized Josephson current and spin-flipper after interaction is in a superposition of mutual spin-flip as well as no flip state given by the joint entangled wave-function of spin-polarized Josephson current and spin-flipper as-
\begin{equation}
\ket{s.c}\otimes\ket{\phi_{m'}^{S}}=\sqrt{\frac{f_{2}}{2}}\ket{\mbox{Mutual-Flip}}+\sqrt{\frac{m'}{2}}\ket{\mbox{No flip}}
\end{equation}
where Josephson current state $|s.c\rangle$ is spin polarized. Quantum spin flip scattering plays an integral role in observing anomalous Josephson effect as we discuss later. In absence of spin flip scattering probability ($f_{2}=0$), anomalous Josephson current vanishes.

{\textbf{Explaining chiral symmetry breaking-}}
All standard Josephson junctions have a certain symmetry, called chiral symmetry. Due to this symmetry, at $\varphi=0$ one cannot distinguish between electron tunneling from left to right superconductor and vice-versa. Thus, electron tunneling amplitude from left to right superconductor equals that from right to left superconductor when there is no phase difference between two superconductors ($\varphi=0$). Thus for our system, as in Fig.~1, when Ferromagnet's are aligned ($\theta=0$), $t_{ee}^{\uparrow\uparrow}(\varphi=0)\mid_{L\rightarrow R}=t_{ee}^{\uparrow\uparrow}(\varphi=0)\mid_{L\leftarrow R}$, where $t_{ee}^{\uparrow\uparrow}\mid_{L\rightarrow R}$ and $t_{ee}^{\uparrow\uparrow}\mid_{L\leftarrow R}$ are the transmission amplitudes for electron tunneling from left to right superconductor and vice-versa. This implies chiral symmetry is not broken and as a result, $I(\varphi=0)$ is strictly zero. But, when Ferromagnet's are misaligned ($\theta\neq0$), then $t_{ee}^{\uparrow\uparrow}(\varphi=0)\mid_{L\rightarrow R}\neq t_{ee}^{\uparrow\uparrow}(\varphi=0)\mid_{L\leftarrow R}$, i.e., chiral symmetry is broken.

{\textbf{Explaining time reversal symmetry breaking-}}
Hamiltonian matrix, in Eq.~(1) is denoted as $H_{BdG}(\varphi)$, such that $H_{BdG}(\varphi)\psi(x)=E\psi(x)$. When spin flip probability $f_{2}=0$ or Ferromagnet's are aligned $\theta=0$, $H_{BdG}(\varphi)$ preserves time reversal symmetry ($T$), thus $TH_{BdG}(\varphi)T^{\dagger}=H_{BdG}(-\varphi)$, which implies that if $H_{BdG}(\varphi)$ possess an energy eigenvalue $\varepsilon_{l}(\varphi)$, then $H_{BdG}(-\varphi)$ must have the same energy eigenvalue. The Andreev bound states then satisfy: $\varepsilon_{l}(\varphi)=\varepsilon_{l}(-\varphi)$, and as a result for Josephson current $I(\varphi)=-I(-\varphi)$ and there is no anomalous Josephson effect. In presence of spin flip scattering ($f_{2}\neq0$) and when Ferromagnet's are misaligned ($\theta\neq0$), time reversal symmetry is broken, as a result, $\varepsilon_{l}(\varphi)\neq\varepsilon_{l}(-\varphi)$, i.e., Andreev bound state symmetry is also broken and thus Josephson current obeys $I(-\varphi)\neq -I(\varphi)$, which implies $I(\varphi=0)\neq0$. Thus, when both $f_{2}\neq0$ and $\theta\neq0$, i.e., both time reversal symmetry and chiral symmetry are broken, an anomalous Josephson current flows across the junction. In contrast, when $f_{2}=0$ and $\theta\neq0$, i.e., only chiral symmetry is broken, but time reversal symmetry is not broken then anomalous Josephson current vanishes.
{\subsection{\textbf{How different values of $T/T_{c}$ affect anomalous Josephson current?}}
In Fig.~6, we show the effect of finite temperature on anomalous Josephson current, anomalous phase and asymmetry of the critical current in presence of scattering with $f_{2}=1$, i.e., $S=1/2$ and $m'=-1/2$, while $J=1$ for transparent junction. In Fig.~6(a) of our manuscript, we plot anomalous Josephson current as a function of mis-orientation angle ($\theta$) between two ferromagnetic layers for different values of $T/T_{c}$. We see that magnitude of anomalous current increases with increasing $T/T_{c}$. Further, sign of anomalous Josephson current does not change with $T/T_{c}$. 
In Fig.~6(b) we plot asymmetry of the critical current $\aleph$ as a function of exchange interaction $J$ for different values of $T/T_{c}$. We see that the maximum value of $\aleph$ increases with increasing temperature. In Fig.~6(c) we plot anomalous phase $\varphi_{0}$ as a function of exchange interaction $J$ of spin flipper for different values of $T/T_{c}$. We see that magnitude of anomalous phase $\varphi_{0}$ increases with increasing $T/T_{c}$ although magnitude of the \textbf{\textquotedblleft quantized\textquotedblright} steps in the anomalous phase $\varphi_{0}$ remain unchanged, i.e., $\pi/100$ radians regardless of $T/T_{c}$, meaning the quantization of steps at the value $\pi/100$ radians is independent of $T/T_{c}$. Further this quantization at $\pi/100$ radians is independent of $J$, $Z$, $h$, $\theta$, $S$, $m'$ and $k_{F}a$ suggesting this is an universal feature in our device.}
\section{Experimental realization and Conclusions}
The set-up as envisaged in Fig.~1 can be realized in a experimental lab. Superconductor-Ferromagnet-Ferromagnet-Superconductor (S-F-F-S) Josephson junctions have been designed experimentally for quite some time now\cite{exp}. Embedding a S-F-F-S junction with a magnetic adatom or spin-flipper at the interface between two ferromagnets shouldn't be difficult, especially with an s-wave superconductor like Aluminum or Lead it should be perfectly possible. In Ref.~\cite{yaz}, local electronic properties of the surface of a superconductor are studied experimentally in the vicinity of a magnetic adatom with a scanning tunneling microscope (STM). Further, in Ref.~\cite{ruby}, iron (Fe) chains are doped on the superconducting Pb surface and the subgap spectra is examined using scanning tunneling microscope. {In Table 1 we discuss the different properties like time reversal symmetry, chiral symmetry, anomalous Josephson current and Josephson current for three distinct cases: (a) finite spin flip scattering but Ferromagnets are aligned, i.e., $f_{2}\neq0$ but $\theta=0$, (b) no spin flip scattering but Ferromagnets are misaligned, i.e., $f_{2}=0$ but $\theta\neq0$ and (c) when spin flip scattering is finite and Ferromagnets are misaligned, i.e., $f_{2}\neq0$ and $\theta\neq0$. We see that when $f_{2}\neq0$ and $\theta=0$, both time reversal symmetry and chiral symmetry are preserved, as a result anomalous Josephson current vanishes and Josephson current satisfies the relation $I(\varphi)=-I(-\varphi)$. For $f_{2}=0$ and $\theta\neq0$, chiral symmetry is broken but time reversal symmetry is preserved, as a result again anomalous Josephson current is zero and Josephson current follows $I(\varphi)=-I(-\varphi)$. In contrast, when $f_{2}\neq0$ and $\theta\neq0$, both time reversal symmetry and chiral symmetry are broken, as a result anomalous Josephson current flows through the junction and Josephson current obeys $I(\varphi)\neq-I(-\varphi)$.} To conclude, we have studied anomalous Josephson effect and the direction dependent critical current in S-$F_{1}$-spin flipper-$F_{2}$-S junction where $F_{1}$, $F_{2}$ are the two ferromagnetic layers with misaligned magnetization. In absence of spin flip scattering, Andreev bound states are time reversal symmetric, i.e., $\varepsilon_{l}(\varphi)=\varepsilon_{l}(-\varphi)$. As a result, Josephson current is sinusoidal with $I(\varphi)=-I(-\varphi)$ and there is no anomalous Josephson supercurrent at $\varphi=0$. {But in presence of spin flip scattering, anomalous Josephson effect is seen. Andreev bound states break time reversal symmetry, i.e., $\varepsilon_{l}(\varphi)\neq\varepsilon_{l}(-\varphi)$ as well as chiral symmetry, as a result, Josephson current breaks phase inversion symmetry $I(\varphi)\neq -I(-\varphi)$, and an anomalous Josephson current can flow at phase difference $\varphi=0$.} 
\begin{table*}[t]
\centering
\caption{Effect of breaking chiral and/or time reversal symmetry on anomalous Josephson current (spin flip probability is $f_{2}$; misorientation angle=$\theta$}  
\begin{tabular}{ |p{2.95cm}|p{4.45cm}|p{4.45cm}|p{4.45cm}|} 
\hline 
Parameters$\rightarrow$ Properties$\downarrow$ & $f_{2}\neq0$, $\theta=0$ & $f_{2}=0$, $\theta\neq0$ & $f_{2}\neq0$ and $\theta\neq0$\\
\hline
\scriptsize{Time reversal symmetry} & Preserved, $\varepsilon_{l}(\varphi)=\varepsilon_{l}(-\varphi)$ & Preserved, $\varepsilon_{l}(\varphi)=\varepsilon_{l}(-\varphi)$ & Broken, $\varepsilon_{l}(\varphi)\neq\varepsilon_{l}(-\varphi)$\\
\hline
\scriptsize{Chiral symmetry} & Preserved, & Broken, & Broken,\\
& \scriptsize{$t_{ee}^{\uparrow\uparrow}(\varphi=0)\mid_{L\rightarrow R}=t_{ee}^{\uparrow\uparrow}(\varphi=0)\mid_{L\leftarrow R}$} & \scriptsize{$t_{ee}^{\uparrow\uparrow}(\varphi=0)\mid_{L\rightarrow R}\neq t_{ee}^{\uparrow\uparrow}(\varphi=0)\mid_{L\leftarrow R}$} & \scriptsize{$t_{ee}^{\uparrow\uparrow}(\varphi=0)\mid_{L\rightarrow R}\neq t_{ee}^{\uparrow\uparrow}(\varphi=0)\mid_{L\leftarrow R}$}\\
\hline
\scriptsize{Anomalous Josephson current} & Zero & Zero & Finite\\
\hline
\scriptsize{Josephson current} & $I(-\varphi)=-I(\varphi)$ & $I(-\varphi)=-I(\varphi)$ & $I(-\varphi)\neq-I(\varphi)$\\ 
\hline
\end{tabular}
\end{table*}
Further, our system acts as a phase battery which can store quantized amounts of anomalous phase $\varphi_{0}$ in the ground state of the junction.
\acknowledgments
This work was supported by the SERB grant EMR/2015/001836:``Non-local correlations in nanoscale systems: Role of decoherence, interactions, disorder and pairing symmetry'', Principal Investigator: Dr. C. Benjamin.

\section{Supplementary Material}
In this supplementary material, we first outline a brief sketch of our set-up and give a theoretical background to our study by writing Hamiltonian, wavefunctions and boundary conditions to calculate anomalous Josephson current and anomalous phase. In Appendix section we also provide explicit form of expression of anomalous Josephson current and show that for no flip case or absence of spin-flipper or when magnetizations of Ferromagnet's are aligned, anomalous Josephson current vanishes.
\section{Hamiltonian, Wavefunctions and boundary conditions in the Ferromagnetic Josephson junction in presence of a spin flipper}
\subsection{\textbf{Hamiltonian}}
\begin{figure}[ht]
\includegraphics[width=0.48\textwidth]{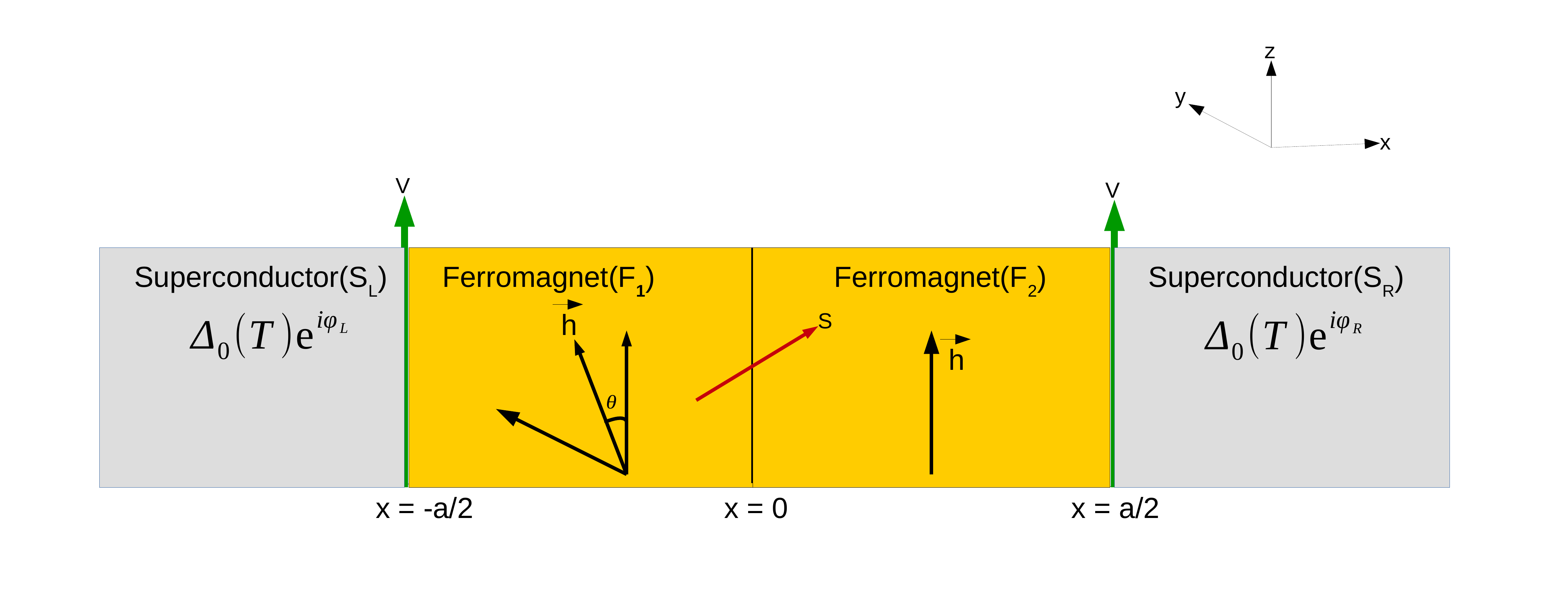}
\caption{\small \sl Josephson junction with two Ferromagnet's and a spin-flipper (spin $S$, magnetic moment $m'$) at $x=0$ sandwiched between two s-wave superconductors.}
\label{AJ}
\end{figure}

We consider a system which consists of two Ferromagnet's ($F_{1}$ and $F_{2}$) with a spin-flipper, embedded between two s-wave singlet superconductors. Our set-up is depicted in Fig.~\ref{AJ}, it shows a spin-flipper at $x=0$ and two superconductors- one to left $x<-a/2$ and another at right $x>a/2$. There are two Ferromagnet's in between at $-a/2<x<0$ and $0<x<a/2$. The magnetization vectors of the two Ferromagnet's make an angle $\theta$ with each other. We take the superconducting gap of the form $\Delta=\Delta_{0}(T)[e^{i\varphi_{L}}\Theta(-x-a/2)+e^{i\varphi_{R}}\Theta(x-a/2)]$, where $\Delta_{0}(T)$ is temperature dependent and it follows that $\Delta_{0}(T)=\Delta_{0}\tanh(1.74\sqrt{(T_{c}/T-1)})$, where $T_{c}$ is the superconducting critical temperature\cite{annu}, $\varphi_{L}$ and $\varphi_{R}$ being superconducting phases for left and right superconductors respectively.

The Hamiltonian in Bogoliubov-de Gennes formalism for our system is\cite{LINDER}-
\begin{eqnarray}
\begin{pmatrix}
H\hat{I} & i\Delta \hat{\sigma}_{y}\\
-i\Delta^{*} \hat{\sigma}_{y} & -H^{*}\hat{I}
\end{pmatrix} \psi(x)& =& E \psi(x), \mbox{}
\label{eqq}
\end{eqnarray}
with $H=p^2/2m^{\star}+V[\delta(x+a/2)+\delta(x-a/2)]-J_{0}\delta(x)\vec s.\vec S-\vec{h}.\hat{\sigma}[\Theta(x+a/2)+\Theta(a/2-x)]-E_{F}$, with $p^2/2m^{\star}$ being the kinetic energy of electron with mass $m^{\star}$, $V$ denotes the strength of $\delta$ potentials at the interfaces between Ferromagnet's and Superconductor, $J_{0}$ denotes strength of exchange coupling between electron/hole with spin $\vec{s}$ and spin-flipper\cite{joseph} with spin $\vec{S}$. $\psi(x)$ defines a four-component spinor, while $E_{F}$ is the Fermi energy, $\hat{\sigma}$'s are Pauli spin matrices and $\hat{I}$ is $2\times2$ identity matrix. The magnetization vector ($\vec{h}$) of left ferromagnetic layer ($F_{1}$) is at an angle $\theta$ with $z$ axis in the $y-z$ plane, while that of right ferromagnetic layer ($F_{2}$) is fixed along the $z$ axis. Thus, $\vec{h}.\hat{\sigma}=h\sin \theta\hat{\sigma}_{y}+h\cos \theta\hat{\sigma}_{z}$\cite{Halter}.
\subsection{\textbf{Wave-functions}}
If a spin up electron is incident at the $x=-a/2$ interface from left superconductor. Then wave function in the left superconductor ($S_{L}$) is given by\cite{LINDER}-
\begin{align}
\begin{split}
\psi_{S_{L}}(x)={}&\begin{pmatrix}u\\
                              0\\
                              0\\ 
                              v
                              \end{pmatrix}e^{ik_{+}x}\phi_{m'}^{S}+r_{ee}^{\uparrow\uparrow}\begin{pmatrix}
                              u\\
                              0\\
                              0\\
                              v
                             \end{pmatrix}e^{-ik_{+}x}\phi_{m'}^{S}+ r_{ee}^{\uparrow\downarrow}\begin{pmatrix}
                             0\\
                             u\\
                            -v\\
                             0
                             \end{pmatrix}\\{}&e^{-ik_{+}x}\phi_{m'+1}^{S}+r_{eh}^{\uparrow\uparrow}\begin{pmatrix}
                             0\\
                            -v\\
                             u\\
                             0
                             \end{pmatrix}e^{ik_{-}x}\phi_{m'+1}^{S}+r_{eh}^{\uparrow\downarrow}\begin{pmatrix}
                             v\\
                             0\\
                             0\\
                             u
                             \end{pmatrix}\\{}&e^{ik_{-}x}\phi_{m'}^{S},\\
                             \end{split}
                             \end{align}
where $r_{ee}^{\uparrow\uparrow},r_{ee}^{\uparrow\downarrow},r_{eh}^{\uparrow\uparrow},r_{eh}^{\uparrow\downarrow}$ are the amplitudes for normal reflection without flip, normal reflection with spin flip, Andreev reflection with spin flip and Andreev reflection without flip respectively.
The corresponding wave function in the right superconductor ($S_{R}$) is given by-
\begin{align}
\begin{split}
\psi_{S_{R}}(x)={}&t_{ee}^{\uparrow\uparrow}\begin{pmatrix}
                              ue^{i\varphi}\\
                              0\\
                              0\\
                              v
                             \end{pmatrix}e^{ik_{+}x}\phi_{m'}^{S}+t_{ee}^{\uparrow\downarrow}\begin{pmatrix}
                             0\\
                             ue^{i\varphi}\\
                             -v\\
                             0
                             \end{pmatrix}e^{ik_{+}x}\phi_{m'+1}^{S}\\{}&+t_{eh}^{\uparrow\uparrow}\begin{pmatrix}
                             0\\
                             -ve^{i\varphi}\\
                             u\\
                             0
                             \end{pmatrix}e^{-ik_{-}x}\phi_{m'+1}^{S}+t_{eh}^{\uparrow\downarrow}\begin{pmatrix}
                             ve^{i\varphi}\\
                             0\\
                             0\\
                             u
                             \end{pmatrix}\\{}&e^{-ik_{-}x}\phi_{m'}^{S},\\
                             \end{split}
                             \end{align}
where $t_{ee}^{\uparrow\uparrow},t_{ee}^{\uparrow\downarrow},t_{eh}^{\uparrow\uparrow},t_{eh}^{\uparrow\downarrow}$ are the transmission amplitudes, corresponding to the reflection process described above. $\varphi=\varphi_{R}-\varphi_{L}$ represents the phase difference between right and left superconductors. $\phi_{m'}^{S}$ is the eigen-spinor of the spin-flipper, with its $S_{z}$ operator acting as-
$S_{z}\phi_{m'}^{S} = m'\phi_{m'}^{S}$, with $m'$ being the spin magnetic moment in $z-$ direction for the spin flipper. $u=\sqrt{\frac{1}{2}\Bigg(1+\frac{\sqrt{E^2-|\Delta|^{2}}}{E}\Bigg)}$ and $v=\sqrt{\frac{1}{2}\Bigg(1-\frac{\sqrt{E^2-|\Delta|^{2}}}{E}\Bigg)}$ are the BCS coherence factors.
$k_{\pm}=\sqrt{\frac{2m^{\star}}{\hbar^2}(E_{F}\pm \sqrt{E^2-|\Delta|^2})}$ is the wave-vector for electron-like quasi-particle ($k_{+}$) and hole-like quasi-particle ($k_{-}$) in the left and right superconducting wave-functions, $\psi_{S_{L}}$ and $\psi_{S_{R}}$ respectively.\\
The wave-function in the left Ferromagnet ($F_{1}$) is given by-
\begin{align}
\begin{split}
\psi_{F_{1}}(x)={}&(ee^{iq_{\uparrow}^{+}(x+a/2)}+fe^{-iq_{\uparrow}^{+}x})\begin{pmatrix}
                                                     \cos \frac{\theta}{2}\\
                                                     i\sin\frac{\theta}{2}\\
                                                     0\\
                                                     0
                                                    \end{pmatrix}\phi_{m'}^{S}+\\{}&(e^{\prime} e^{iq_{\downarrow}^{+}(x+a/2)}+f^{\prime}e^{-iq_{\downarrow}^{+}x})\begin{pmatrix}
                                                     i\sin \frac{\theta}{2}\\
                                                     \cos\frac{\theta}{2}\\
                                                     0\\
                                                     0
                                                     \end{pmatrix}\phi_{m'+1}^{S}+\\{}&(e_{0}e^{-iq_{\uparrow}^{-}(x+a/2)}+f_{0}e^{iq_{\uparrow}^{-}x})\begin{pmatrix}
                                                     0\\
                                                     0\\
                                                     \cos\frac{\theta}{2}\\
                                                     -i\sin\frac{\theta}{2}
                                                     \end{pmatrix}\phi_{m'+1}^{S}+\\{}&(e_{0}^{\prime} e^{-iq_{\downarrow}^{-}(x+a/2)}+f_{0}^{\prime} e^{iq_{\downarrow}^{-}x})\begin{pmatrix}
                                                     0\\
                                                     0\\
                                                     -i\sin\frac{\theta}{2}\\
                                                     \cos\frac{\theta}{2}
                                                     \end{pmatrix}\phi_{m'}^{S}.\\
                                                    \end{split}
                                                    \end{align}
Similarly the wave-function in the right Ferromagnet ($F_{2}$) is given by-
\begin{align}
\begin{split}
\psi_{F_{2}}(x)={}&(a_{0}e^{iq_{\uparrow}^{+}x}+be^{-iq_{\uparrow}^{+}(x-a/2)})\begin{pmatrix}
                                                     1\\
                                                     0\\
                                                     0\\
                                                     0
                                                    \end{pmatrix}\phi_{m'}^{S}+(a^{\prime} e^{iq_{\downarrow}^{+}x}+\\{}&b^{\prime}
                                                    e^{-iq_{\downarrow}^{+}(x-a/2)})\begin{pmatrix}
                                                     0\\
                                                     1\\
                                                     0\\
                                                     0
                                                     \end{pmatrix}\phi_{m'+1}^{S}+(ce^{-iq_{\uparrow}^{-}x}+d\\{}&e^{iq_{\uparrow}^{-}(x-a/2)}) \begin{pmatrix}
                                                     0\\
                                                     0\\
                                                     1\\
                                                     0
                                                     \end{pmatrix}
                                                    \phi_{m'+1}^{S}+(c^{\prime} e^{-iq_{\downarrow}^{-}x}+d^{\prime}\\{}& e^{iq_{\downarrow}^{-}(x-a/2)})\begin{pmatrix}
                                                     0\\
                                                     0\\
                                                     0\\
                                                     1
                                                     \end{pmatrix}\phi_{m'}^{S},
                                                     \end{split}
                                                    \end{align}
and $q_{\sigma}^{\pm}=\sqrt{\frac{2m^{\star}}{\hbar^2}(E_{F}+\rho_{\sigma}h\pm E)}$ is the wave-vector for electron ($q_{\sigma}^{+}$) and hole ($q_{\sigma}^{-}$) in the Ferromagnet, with $\rho_{\sigma}=+1(-1)$ when- $\sigma=\uparrow(\downarrow)$. In our work we have used the Andreev approximation: $k_{+}=k_{-}=\sqrt{\frac{2m^{\star}E_{F}}{\hbar^2}}=k_{F}$ and $q_{\uparrow,\downarrow}=k_{F}\sqrt{1\pm\frac{h}{E_{F}}}$, where $k_{F}$ is the Fermi wave-vector, with $E_{F}>>|\Delta|, E$.
\subsection{\textbf{Boundary conditions}}
The boundary conditions at $x=-a/2$ is- $\psi_{S_{L}}(x)=\psi_{F_{1}}(x)$ (continuity of wave-functions) and, $\frac{d\psi_{F_{1}}}{dx}-\frac{d\psi_{S_{L}}}{dx}=\frac{2m^{\star}V}{\hbar^2}\psi_{F_{1}}$ (discontinuity in first derivative), and at $x=0$ similarly is\cite{Fra}- $\psi_{F_{1}}(x)=\psi_{F_{2}}(x)$ and $\frac{d\psi_{F_{2}}}{dx}-\frac{d\psi_{F_{1}}}{dx}=-\frac{2m^{\star}J_{0}\vec s.\vec S}{\hbar^2} \psi_{F_{1}}$ where $\vec s.\vec S=s_{z}S_{z}+\frac{1}{2}(s^{-}S^{+}+s^{+}S^{-})$, $s^{\pm} = s_{x}\pm is_{y}$ and $S^{\pm}=S_{x}\pm iS_{y}$, with $s_{k}=\frac{\hbar}{2}\sigma_{k}$, $k=x,y,z$ are the exchange operator due to spin flipper in the Hamiltonian, the spin raising and lowering operators for electron/hole and spin flipper respectively with-
\begin{equation}
\vec s.\vec S\begin{pmatrix}
               1\\
               0\\
               0\\
               0
              \end{pmatrix}\phi_{m'}^{S}=\frac{m'}{2}\begin{pmatrix}
              1\\
              0\\
              0\\
              0
             \end{pmatrix}\phi_{m'}^{S}+\frac{f_{2}}{2}\begin{pmatrix}
             0\\
             1\\
             0\\
             0
             \end{pmatrix}\phi_{m'+1}^{S}, \mbox{ and }
             \end{equation}
\begin{equation}
 \vec s.\vec S\begin{pmatrix}
              0\\
              1\\
              0\\
              0\\
             \end{pmatrix}\phi_{m'+1}^{S}=-\frac{(m'+1)}{2}\begin{pmatrix}
             0\\
             1\\
             0\\
             0
             \end{pmatrix}\phi_{m'+1}^{S}+\frac{f_{2}}{2}\begin{pmatrix}
                                                                1\\
                                                                0\\
                                                                0\\
                                                                0
                                                               \end{pmatrix}\phi_{m'}^{S}.
\end{equation}                                                               
Here, $\vec{s}$ represents spin operator acting on electron/hole states, while $\vec{S}$ represents the spin operator acting on spin flipper states and $f_{2}=\sqrt{(S-m')(S+m'+1)}$ is the spin-flip probability\cite{AJP} for spin flipper.

On the other hand when there is no spin-flip scattering, i.e., if $S=m'$, then spin flip probability of spin flipper: $f_{2}=\sqrt{(S-m')(S+m'+1)}=0$.
\begin{equation}
\mbox{ Thus, for no-flip process: }\vec s.\vec S\begin{pmatrix}
               1\\
               0\\
               0\\
               0
              \end{pmatrix}\phi_{m'}^{S}=\frac{m'}{2}\begin{pmatrix}
              1\\
              0\\
              0\\
              0
             \end{pmatrix}\phi_{m'}^{S}.
             \end{equation}
The Hamiltonian therefore for no-flip process then is- $H_{0}=p^2/2m^{\star}+V[\delta(x+a/2)+\delta(x-a/2)]-J_{0}\delta(x)s^{z}S^{z}-\vec{h}.\hat{\sigma}[\Theta(x+a/2)+\Theta(a/2-x)]-E_{F}$, while flip Hamiltonian for spin flip process is-
$H_{0}=p^2/2m^{\star}+V[\delta(x+a/2)+\delta(x-a/2)]-J_{0}\delta(x)\vec s.\vec S-\vec{h}.\hat{\sigma}[\Theta(x+a/2)+\Theta(a/2-x)]-E_{F}$.             

Finally, at $x=a/2$, the boundary conditions are- $\psi_{F_{2}}(x)=\psi_{S_{R}}(x)$, $\frac{d\psi_{S_{R}}}{dx}-\frac{d\psi_{F_{2}}}{dx}=\frac{2m^{\star}V}{\hbar^2}\psi_{F_{2}}$.
We use the dimensionless parameters
$J=\frac{m^{\star}J_{0}}{\hbar^2k_{F}}$ as a measure of strength of exchange interaction and $Z=\frac{m^{\star}V}{\hbar^2 k_{F}}$ as a measure of interface transparency\cite{BTK}.

\subsection{\textbf{Josephson current}}
To calculate bound state contribution to Josephson current we follow the procedure established in Ref.~\cite{annu}. We neglect the contribution from incoming quasi-particle, i.e., first term $\begin{pmatrix}u & 0 & 0 & v\end{pmatrix}^{T}e^{ik_{+}x}\phi_{m'}^{S}$ of Eq.~(7) and insert the wave-functions into boundary conditions defined in section I. C., we get a homogeneous system of linear equations for the scattering amplitudes,
\begin{equation}
Rx=0, 
\end{equation}
where $x$ is a $8\times1$ column vector, given by $x=\begin{pmatrix}r_{ee}^{\uparrow\uparrow} & r_{ee}^{\uparrow\downarrow} & r_{eh}^{\uparrow\uparrow} & r_{eh}^{\uparrow\downarrow} & t_{ee}^{\uparrow\uparrow} & t_{ee}^{\uparrow\downarrow} & t_{eh}^{\uparrow\uparrow} & t_{eh}^{\uparrow\downarrow}\end{pmatrix}^{T}$, $R$ is a $8\times 8$ matrix obtained by eliminating the scattering amplitudes for the two Ferromagnet's via the scattering amplitudes of the left and right superconductor. For a nontrivial solution of this system, the determinant of $R=0$. Thus, we get the Andreev bound state energy spectrum $E_{i}$, $i=\{1,...,8\}$\cite{Been}. We find that the Andreev energy bound states $E_{i}(i=1,...,8)$ can be written as $\varepsilon_{l}^{\pm}=\pm\varepsilon_{l}(l=1,...,4)$. From Andreev bound state energies we get Free energy of our system, given as-\cite{annu}
\begin{equation}
F=-\frac{1}{\beta}\ln\Big[\prod_{i}(1+e^{-\beta E_{i}})\Big]= -\frac{2}{\beta}\sum_{l=1}^{4}\ln\Big[2\cosh\Big(\frac{\beta \varepsilon_{l}}{2}\Big)\Big]
\label{ff}
\end{equation}
We consider only the short junction limit, such that the total Josephson current can be determined by considering the bound state contribution only. The Josephson current at finite temperature is defined as the derivative of the Free energy $F$ of our system with respect to the phase difference $\varphi$ between left and right superconductors\cite{deG},
\begin{equation}
I=\frac{2e}{\hbar}\frac{\partial F}{\partial \varphi}=-\frac{2e}{\hbar}\sum_{l=1}^{4}\tanh\Big(\frac{\beta \varepsilon_{l}}{2}\Big)\frac{\partial \varepsilon_{l}}{\partial \varphi},
\label{ff1}
\end{equation}
herein $e$ is the charge of electron. Eq.~\ref{ff1} is the main working formula of our paper. Using Eq.~\ref{ff1} we can calculate the anomalous Josephson current, which is given as- $I_{an}=I(\varphi=0)$. In case interfaces are completely  transparent, i.e., $Z=0$, we have-
\begin{align}
\begin{split}
I_{an}={}&\frac{2e\Delta_{0}^{2}(T)}{\hbar}\Big(\tanh\Big(\frac{\beta A_{1}}{2}\Big){A_{1}^{\prime}}+\tanh\Big(\frac{\beta A_{2}}{2}\Big){A_{2}^{\prime}}\\{}&+\tanh\Big(\frac{\beta A_{3}}{2}\Big){A_{3}^{\prime}}+\tanh\Big(\frac{\beta A_{4}}{2}\Big){A_{4}^{\prime}}\Big)
\label{an1}
\end{split}
\end{align}
wherein $A_{1}$, $A_{2}$, $A_{3}$, $A_{4}$, $A_{1}^{\prime}$, $A_{2}^{\prime}$, $A_{3}^{\prime}$ and $A_{4}^{\prime}$ are large expressions that depend on exchange interaction ($J$), magnetization of the Ferromagnet's, spin ($S$) and magnetic moment ($m'$) of spin-flipper, phase ($k_{F}a$) accumulated in Ferromagnet's and spin-flip probability of spin-flipper ($f_{2}$). The explicit forms for $A_{k}$'s and $A_{k}^{\prime}$'s ($k=1,2,3,4$) are given in Appendix. In Appendix we show that for no flip ($f_{2}=0$) or absence of spin-flipper ($J=0$) or $\theta=0$ (magnetizations of Ferromagnet's are aligned), anomalous Josephson current vanishes.

For $\varphi_{0}$ Josephson junction the ground state of the junction is at $\varphi=\varphi_{0}$ ($\neq0$ or $\pi$). At $\varphi=\varphi_{0}$, the free energy of the junction is a minimum. By determining this minimum value of the free energy, one can calculate the anomalous phase $\varphi_{0}$ numerically.

\begin{SCfigure}[][h]
\includegraphics[width=0.3\textwidth]{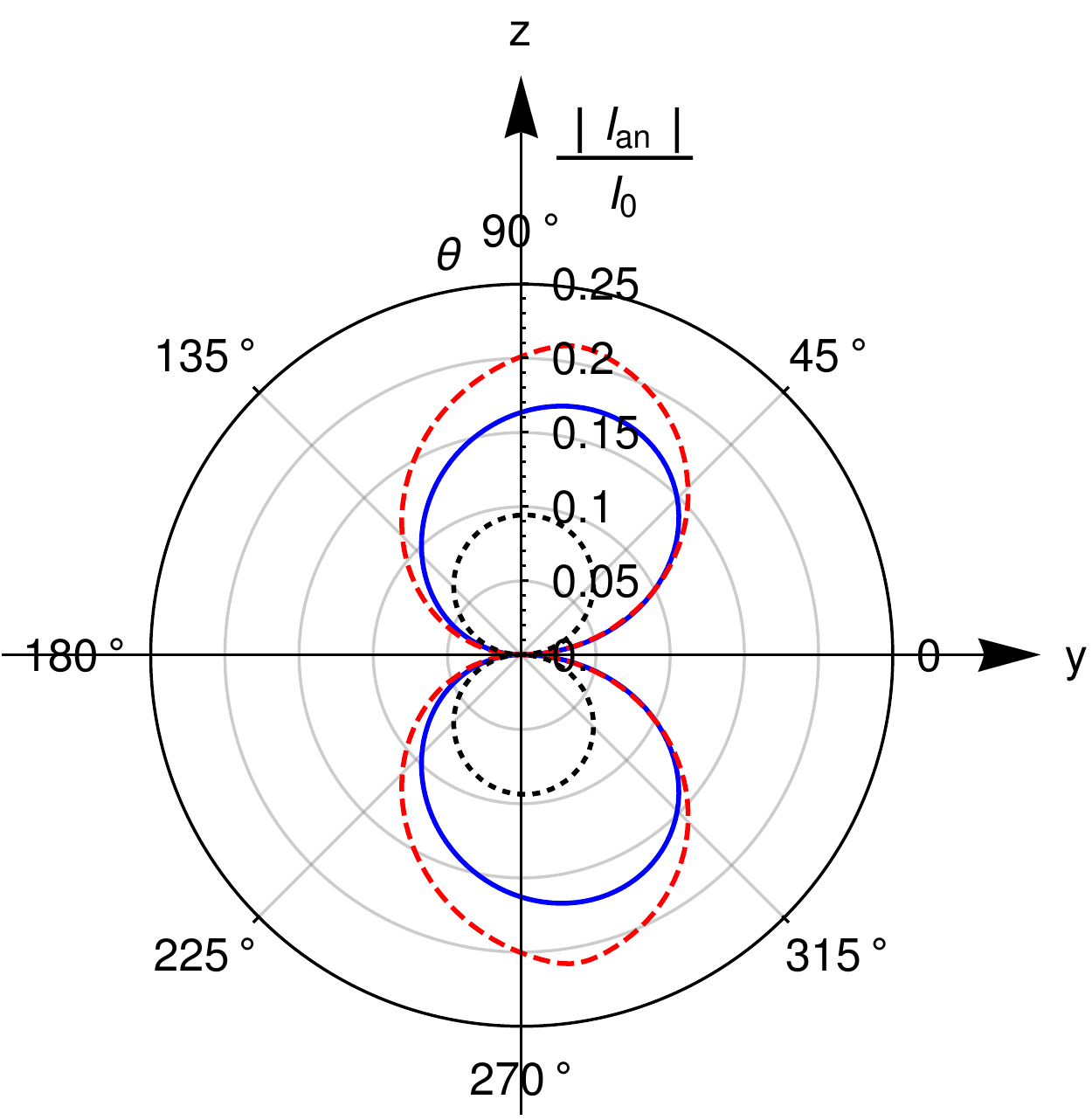}
\caption{ \small \sl Absolute value of anomalous Josephson current as function of the mis-orientation angle ($\theta$) between two Ferromagnet's for different values of spin and magnetic moment of the spin flipper. Parameters are $\Delta_{0}=1meV$, $I_{0}=e\Delta_{0}/\hbar$, $T/T_{c}=0.01$, $J=1$, $h=0.5E_{F}$, $k_{F}a=\pi$, $\varphi=0$, $Z=0$.}
\label{aj1}
\end{SCfigure}

{In Fig.~\ref{aj1} we plot absolute value of the anomalous Josephson currents as function of the mis-orientation angle $\theta$ for same values as in Fig.~3(d) of our main article. We see that the anomalous current regardless of flip probability is symmetric about $y-$axis but asymmetric with respect to $z-$axis.}

\begin{figure}[h]
\includegraphics[width=0.49\textwidth]{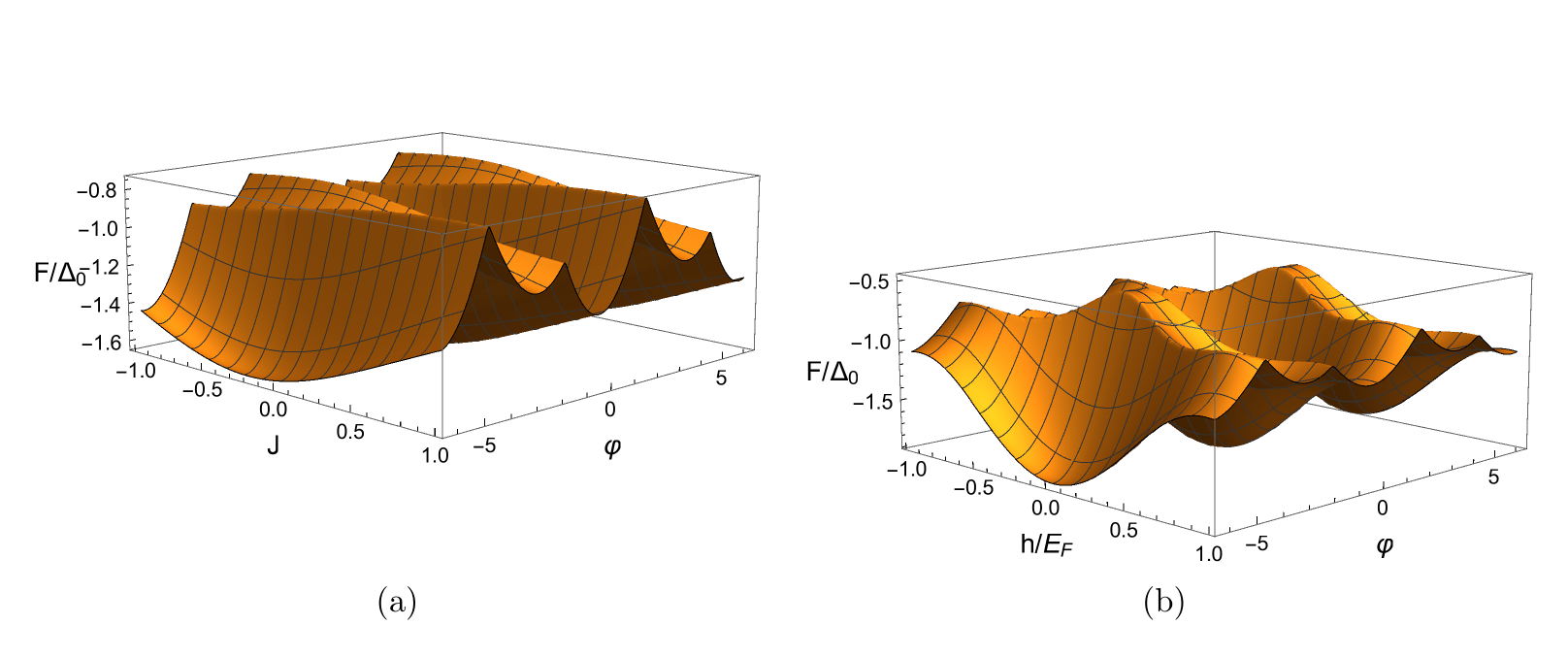}
\caption{ \small \sl (a) Free energy as a function of exchange interaction $J$ of spin flipper and phase difference $\varphi$ across two superconductors. (b) Free energy as a function of magnetization $h$ of the Ferromagnet's and phase difference $\varphi$ across two superconductors. Parameters are $\Delta_{0}=1meV$, $T/T_{c}=0.01$, $S=1/2$, $m'=-1/2$, $h=0.5E_{F}$ (for (a)), $k_{F}a=\pi$, $\theta=\pi/2$, $Z=0$, $J=0.5$ (for (b)).}
\label{aj2}
\end{figure}

In Fig.~\ref{aj2}(a) we plot Free energy as a function of exchange interaction $J$ and phase difference $\varphi$. 
When $J\neq0$, for each particular value of $J$, the minimum Free energy is at $\varphi=\varphi_{0} (\neq0,\pm\pi)$ and a $\varphi_{0}$-Josephson junction is realized\cite{buz,buzd,tan}. Thus, for each particular value of $J$, we get the anomalous phase $\varphi_{0}$ numerically, where the Free energy of the junction becomes minimum. In Fig.~\ref{aj2}(b) we plot Free energy as a function of magnetization $h$ and phase difference $\varphi$.  When $h\neq0$, for each particular value of $h$, the minimum Free energy is at $\varphi=\varphi_{0} (\neq0,\pm\pi)$. Thus, again for each value of $h$, we get an anomalous phase $\varphi_{0}$ numerically, where Free energy of the junction is minimum. This procedure of calculating $\varphi_{0}$ is well known and is also done in Refs.~[\cite{yoko}, \cite{FO}].
\section{Appendix: Explicit form of anomalous Josephson current}
The explicit form of $A_{1}$, $A_{2}$, $A_{3}$, $A_{4}$, $A_{1}^{\prime}$, $A_{2}^{\prime}$, $A_{3}^{\prime}$, $A_{4}^{\prime}$ in Eq.~\ref{an1} is
\scriptsize{
\begin{eqnarray}
A_{1(2)}=&&\Delta_{0}(T)\sqrt{K-\frac{1}{2}\sqrt{L+M}\pm\frac{1}{2}\sqrt{2L-M-\frac{2N}{\sqrt{L+M}}}},\nonumber\\ 
A_{1(2)}^{\prime}=&&-\frac{1}{2A_{1(2)}}\Big(-K^{\prime}-\frac{L^{\prime}+M^{\prime}}{4\sqrt{L+M}}\pm\nonumber\\
                  &&\frac{2L^{\prime}-M^{\prime}+\frac{N(L^{\prime}+M^{\prime})}{(L+M)^{3/2}}-\frac{2N^{\prime}}{(L+M)}}{4\sqrt{2L-M-\frac{2N}{\sqrt{L+M}}}}\Big),\nonumber\\
A_{3(4)}=&&\Delta_{0}(T)\sqrt{K+\frac{1}{2}\sqrt{L+M}\pm\frac{1}{2}\sqrt{2L-M+\frac{2N}{\sqrt{L+M}}}},\nonumber\\
\mbox{ and }A_{3(4)}^{\prime}=&&-\frac{1}{2A_{3(4)}}\Big(-K^{\prime}+\frac{L^{\prime}+M^{\prime}}{4\sqrt{L+M}}\pm\nonumber\\
                              &&\frac{2L^{\prime}-M^{\prime}-\frac{N(L^{\prime}+M^{\prime})}{(L+M)^{3/2}}+\frac{2N^{\prime}}{(L+M)}}{4\sqrt{2L-M+\frac{2N}{\sqrt{L+M}}}}\Big),\nonumber
\end{eqnarray}
where
\begin{align}
\begin{split}
L={}&4T_{1}^2-\frac{2}{3}T_{2},\\
M={}&\frac{2^{1/3}X_{1}}{3(X_{2}+\sqrt{X_{2}^2-4X_{1}^3})}+\frac{(X_{2}+\sqrt{X_{2}^2-4X_{1}^3})^{1/3}}{2^{1/3}3},\\
N={}&8T_{1}^{3}-2T_{1}T_{2}+T_{3},\\\nonumber
K={}&T_{1},\\
\end{split}
\end{align}
\begin{align}
\begin{split}
L^{\prime}={}&-8T_{1}U_{1}-\frac{2}{3}U_{2},\\
M^{\prime}={}&\frac{2^{1/3}X_{1}^{\prime}}{3(X_{2}+\sqrt{X_{2}^2-4X_{1}^3})^{1/3}}-\frac{2^{1/3}X_{1}Y}{9(X_{2}+\sqrt{X_{2}^2-4X_{1}^3})^{4/3}},\\
N^{\prime}={}&-192T_{1}^{2}U_{1}+16T_{2}U_{1}-16T_{1}U_{2}+8U_{3},\\
K^{\prime}={}&U_{1},\\
\mbox{ where }\\
X_{1}={}&T_{2}^{2}-12T_{1}T_{3}-12T_{4},\\
X_{2}={}&2T_{2}^{3}-36T_{1}T_{2}T_{3}-432T_{1}^{2}T_{4}+27T_{3}^{2}+72T_{2}T_{4},\\
X_{1}^{\prime}={}&2T_{2}U_{2}+12T_{3}U_{1}-12T_{1}U_{3},\\
Y={}&Y^{\prime}+\frac{X_{2}Y^{\prime}-6X_{1}^{2}X_{1}^{\prime}}{\sqrt{X_{2}^2-4X_{1}^3}},\\
Y^{\prime}={}&6T_{2}^2U_{2}-36T_{1}T_{3}U_{2}+36T_{2}T_{3}U_{1}+864T_{1}T_{4}U_{1}+54T_{3}U_{3}\\{}&+72T_{4}U_{2},\\
T_{1}={}&\Big(P_{1}(S,m',f_{2},h,J,k_{F}a)+P_{2}(S,m',f_{2},h,J,k_{F}a)\cos(\theta)+\\
{}&P_{3}(S,m',f_{2},h,J,k_{F}a)\cos(2\theta)+P_{4}(S,m',f_{2},h,J,k_{F}a)
\cos(3\theta)\\{}&+P_{5}(S,m',f_{2},h,J,k_{F}a)\cos(4\theta)\Big)/\Big(Q_{1}(S,m',f_{2},h,J,k_{F}a)+\\
{}&Q_{2}(S,m',f_{2},h,J,k_{F}a)\cos(\theta)+Q_{3}(S,m',f_{2},h,J,k_{F}a)\cos(2\theta)\\
{}&+Q_{4}(S,m',f_{2},h,J,k_{F}a)\cos(3\theta)+Q_{5}(S,m',f_{2},h,J,k_{F}a)\\{}&\cos(4\theta)\Big),\\
T_{2}={}&\Big(P_{6}(S,m',f_{2},h,J,k_{F}a)+P_{7}(S,m',f_{2},h,J,k_{F}a)\cos(\theta)+\\
      {}&P_{8}(S,m',f_{2},h,J,k_{F}a)\cos(2\theta)+P_{9}(S,m',f_{2},h,J,k_{F}a)\cos(3\theta)\\{}&+P_{10}(S,m',f_{2},h,J,k_{F}a)\cos(4\theta)\Big)/\Big(Q_{1}(S,m',f_{2},h,J,k_{F}a)+\\{}&Q_{2}(S,m',f_{2},h,J,k_{F}a)\cos(\theta)+Q_{3}(S,m',f_{2},h,J,k_{F}a)\cos(2\theta)\\{}&+Q_{4}(S,m',f_{2},h,J,k_{F}a)\cos(3\theta)+Q_{5}(S,m',f_{2},h,J,k_{F}a)\\{}&\cos(4\theta)\Big),\\
T_{3}={}&\Big(P_{11}(S,m',f_{2},h,J,k_{F}a)+P_{12}(S,m',f_{2},h,J,k_{F}a)\cos(\theta)+\\{}&P_{13}(S,m',f_{2},h,J,k_{F}a)\cos(2\theta)+P_{14}(S,m',f_{2},h,J,k_{F}a)\\{}&\cos(3\theta)+P_{15}(S,m',f_{2},h,J,k_{F}a)\cos(4\theta)\Big)/\\{}&\Big(Q_{1}(S,m',f_{2},h,J,k_{F}a)+Q_{2}(S,m',f_{2},h,J,k_{F}a)\cos(\theta)+\\{}&Q_{3}(S,m',f_{2},h,J,k_{F}a)\cos(2\theta)+Q_{4}(S,m',f_{2},h,J,k_{F}a)\cos(3\theta)\\{}&+Q_{5}(S,m',f_{2},h,J,k_{F}a)\cos(4\theta)\Big),\\
T_{4}={}&\Big(P_{16}(S,m',f_{2},h,J,k_{F}a)+P_{17}(S,m',f_{2},h,J,k_{F}a)\cos(\theta)+\\{}&P_{18}(S,m',f_{2},h,J,k_{F}a)\cos(2\theta)+P_{19}(S,m',f_{2},h,J,k_{F}a)
\\{}&\cos(3\theta)+P_{20}(S,m',f_{2},h,J,k_{F}a)\cos(4\theta)\Big)/\\{}&\Big(Q_{1}(S,m',f_{2},h,J,k_{F}a)+Q_{2}(S,m',f_{2},h,J,k_{F}a)\cos(\theta)+\\{}&Q_{3}(S,m',f_{2},h,J,k_{F}a)\cos(2\theta)+Q_{4}(S,m',f_{2},h,J,k_{F}a)\cos(3\theta)\\{}&+Q_{5}(S,m',f_{2},h,J,k_{F}a)\cos(4\theta)\Big),\\
U_{1}={}&\Big(Jf_{2}\sin(\theta)\Big(P_{21}(S,m',f_{2},h,J,k_{F}a)+P_{22}(S,m',f_{2},h,J,k_{F}a)\\{}&\cos(\theta)+P_{23}(S,m',f_{2},h,J,k_{F}a)\cos(2\theta)\Big)\Big)/\\{}&\Big(Q_{6}(S,m',f_{2},h,J,k_{F}a)+Q_{7}(S,m',f_{2},h,J,k_{F}a)\cos(\theta)+\\{}&Q_{8}(S,m',f_{2},h,J,k_{F}a)\cos(2\theta)+Q_{9}(S,m',f_{2},h,J,k_{F}a)\cos(3\theta)\\{}&+Q_{10}(S,m',f_{2},h,J,k_{F}a)\cos(4\theta)\Big),\\
U_{2}={}&\Big(Jf_{2}\sin(\theta)\Big(P_{24}(S,m',f_{2},h,J,k_{F}a)+P_{25}(S,m',f_{2},h,J,k_{F}a)\\{}&\cos(\theta)+P_{26}(S,m',f_{2},h,J,k_{F}a)\cos(2\theta)\Big)\Big)/\\{}&\Big(Q_{1}(S,m',f_{2},h,J,k_{F}a)+Q_{2}(S,m',f_{2},h,J,k_{F}a)\cos(\theta)+\\{}&Q_{3}(S,m',f_{2},h,J,k_{F}a)\cos(2\theta)+Q_{4}(S,m',f_{2},h,J,k_{F}a)\cos(3\theta)\\{}&+Q_{5}(S,m',f_{2},h,J,k_{F}a)\cos(4\theta)\Big),\\\nonumber
\end{split}
\end{align}
\begin{align}
\begin{split}
U_{3}={}&\Big(Jf_{2}\sin(\theta)\Big(P_{27}(S,m',f_{2},h,J,k_{F}a)+P_{28}(S,m',f_{2},h,J,k_{F}a)\\{}&\cos(\theta)+P_{29}(S,m',f_{2},h,J,k_{F}a)\cos(2\theta)\Big)\Big)/\\{}&\Big(Q_{1}(S,m',f_{2},h,J,k_{F}a)+Q_{2}(S,m',f_{2},h,J,k_{F}a)+\cos(\theta)\\{}&Q_{3}(S,m',f_{2},h,J,k_{F}a)\cos(2\theta)+Q_{4}(S,m',f_{2},h,J,k_{F}a)\cos(3\theta)\\{}&+Q_{5}(S,m',f_{2},h,J,k_{F}a)\cos(4\theta)\Big)\nonumber\\
\end{split}
\end{align}}
\small{
Here, $P_{i}$ ($i=1,2...29$) and $Q_{i}$ ($i=1,2,...10$) are functions of all parameters like exchange interaction ($J$), magnetization of the Ferromagnets ($h$), spin ($S$) and magnetic moment ($m'$) of spin flipper, phase ($k_{F}a$) accumulated in ferromagnetic region and spin flip probability of spin flipper ($f_{2}$). Since these are large expressions we do not explicitly write them here. From the above expressions for no flip ($f_{2}=0$) or absence of spin flipper ($J=0$) or $\theta=0$ (magnetizations of the Ferromagnets are aligned), $U_{1}$, $U_{2}$, $U_{3}$ and also $L^{\prime}$, $M^{\prime}$, $N^{\prime}$ and $K^{\prime}$ vanish. Thus, from Eq.~\ref{an1}, $A_{1(2)}^{\prime}=0$ and $A_{3(4)}^{\prime}=0$, implying for no flip case or absence of spin flipper or $\theta=0$ anomalous Josephson current vanishes ($I_{an}=0$).}

\end{thebibliography}
\end{document}